\documentclass[sigconf]{acmart}

\renewcommand\footnotetextcopyrightpermission[1]{}  
\usepackage{xspace}
\usepackage{longtable}
\usepackage{array}
\usepackage[table]{xcolor}
\usepackage{xcolor}

\newcommand{\newtext}[1]{\textcolor{black}{#1}}
\AtBeginDocument{%
  }

\acmConference{}{}{}

\usepackage{draftwatermark}
\SetWatermarkText{Accepted – Pre-camera-ready}
\SetWatermarkScale{0.17}
\SetWatermarkLightness{0.85}

\begin{document}

\title{A Gray Literature Study on Fairness Requirements in \\ AI-enabled Software Engineering}

\author{Thanh Nguyen}
\email{thanh.nguyen2@ucalgary.ca}
\orcid{0000-0003-3235-6530}
\affiliation{%
  \institution{University of Calgary}
  \city{Calgary}
  \state{Alberta}
  \country{Canada}
}
\author{Chaima Boufaied}
\email{chaima.boufaied@ucalgary.ca}
\orcid{0000-0003-3235-6530}
\affiliation{%
  \institution{University of Calgary}
  \city{Calgary}
  \state{Alberta}
  \country{Canada}
}
\author{Ronnie de Souza Santos}
\email{ronnie.desouzasantos@ucalgary.ca}
\orcid{0000-0003-3235-6530}
\affiliation{%
  \institution{University of Calgary}
  \city{Calgary}
  \state{Alberta}
  \country{Canada}
}
\renewcommand{\shortauthors}{Nguyen et al.}
\begin{abstract}
Today, with the growing obsession with applying Artificial Intelligence (AI), particularly Machine Learning (ML), to software across various contexts, much of the focus has been on the effectiveness of AI models, often measured through common metrics such as \emph{F1- score}, while fairness receives relatively little attention. 
This paper presents a review of existing gray literature, examining fairness requirements in AI context, with a focus on how they are defined across various application domains, managed throughout the Software Development Life Cycle (SDLC), and the causes, as well as the corresponding consequences of their violation by AI models. 
Our gray literature investigation shows various definitions of fairness requirements in AI systems, commonly emphasizing non- discrimination and equal treatment across different demographic and social attributes. Fairness requirement management practices vary across the SDLC, particularly in model training and bias mitigation, fairness monitoring and evaluation, and data handling practices. Fairness requirement violations are frequently linked, but not limited, to data representation bias, algorithmic and model design bias, human judgment, and evaluation and transparency gaps. The corresponding consequences include harm in a broad sense, encompassing specific professional and societal impacts as key examples, stereotype reinforcement, data and privacy risks, and loss of trust and legitimacy in AI-supported decisions. These findings emphasize the need for consistent frameworks and practices to integrate fairness into AI software, paying as much attention to fairness as to effectiveness.
\end{abstract}


\maketitle

\newcommand{\searchStrings}{3\xspace}
\newcommand{\topResults}{100\xspace}
\newcommand{\totalResults}{400\xspace}
\newcommand{\initialResults}{100\xspace}
\section{Introduction}
\label{intro}

As Artificial Intelligence (AI) and Machine Learning (ML) become integral to software systems, fairness has emerged as a critical concern throughout the Software Development Life Cycle (SDLC)~\cite{brun2018software, soremekun2022software}. 
Fairness in AI/ML refers to the equitable treatment of individuals or groups~\cite{chen2024fairness,ferrara2024fairness,lu2022towards} by the corresponding AI/ ML models, ensuring that their outcomes do not exhibit any bias and are not influenced by sensitive attributes such as gender, race, or socioeconomic status~\cite{cheng2021socially}. 
When such bias occurs, it can lead to discriminatory results across domains, including preference for wealthy individuals in loan approvals, gender bias in hiring, mostly favoring men over women, unfair criminal risk assessments, and misclassification of dark-skinned females by facial recognition models~\cite{cheng2021socially, chen2024fairness, simpson2024parity, dehal2024exposing}. 
These examples illustrate how AI-based decisions can perpetuate and amplify social inequalities when fairness is not explicitly integrated into the SDLC~\cite{dehal2024exposing, soremekun2022software}.

Bias in AI systems often originates from a misalignment between fairness principles and model behavior, known as fairness bugs~\cite{chen2024fairness}. Addressing these issues requires recognizing fairness as a core system requirement rather than an after-deployment property~\cite{baresi2023understanding}. However, defining fairness requirements is challenging, as fairness lacks a universal definition and depends heavily on the application context~\cite{baresi2023understanding, mehrabi2021survey}. 
Existing definitions, such as statistical parity, equalized odds, and fairness through awareness or unawareness~\cite{chen2024fairness, baresi2023understanding}, capture distinct aspects of fairness but involve trade-offs that prevent their simultaneous satisfaction~\cite{chouldechova2017fair} (a.k.a., the impossibility theorem~\cite{baresi2023understanding}).
This highlights a gap in defining and operationalizing fairness, as existing definitions capture different aspects of fairness without a single, comprehensive approach.

While frameworks like Aequitas~\cite{saleiro2018aequitas}, AI Fairness 360~\cite{bellamy2019ai}, and Fairlearn~\cite{bird2020fairlearn} address fairness through bias detection and mitigation, they focus primarily on model development rather than requirement elicitation. This leaves a gap in ensuring fairness from the earliest stages of system design, indicating that requirements engineering remains an underexplored phase for embedding fairness in AI systems~\cite{baresi2023understanding}.

Given the aforementioned gaps and challenges in defining and integrating fairness early in the SDLC, this study investigates how fairness requirements are described and operationalized in gray literature sources that discuss AI and ML systems. 
\newtext{Our choice of investigating a gray literature is motivated by the fact that the latter provides practical, up-to-date insights such as industry guidelines and reports that are often not found in academic, peer-reviewed papers}.
Our main goal is to understand how fairness requirements are defined, managed throughout the SDLC, as well as to explore the circumstances under which they may be violated and the resulting consequences. Our overarching research question is: \textit{How does the gray literature conceptualize and apply fairness requirements and what causes and effects of fairness violations are reported throughout the AI/ML development lifecycle?}
This work contributes a structured analysis of fairness requirements as represented in non-academic yet influential sources producing the following contributions:
\begin{enumerate}
    \item \textbf{Definition and Conceptualization of Fairness Requirements.}  
    We identify how fairness requirements are defined and conceptualized in trustworthy gray literature, providing insight into how fairness is interpreted in practice.
    \item \textbf{Management of Fairness Requirements across the SDLC.}  
    We describe how fairness requirements are managed throughout the SDLC, clarifying where and how fairness is considered in AI development processes.
    \item \textbf{Causes and Effects of Fairness Requirement Violations.}  
    We analyze the main causes and effects of fairness requirement violations, revealing patterns that influence system reliability and trust.
    \item \textbf{Implications for Software Engineering Practice.}  
    We discuss implications for integrating fairness requirements more effectively into software engineering from research and practice perspectives, supporting more equitable and transparent AI systems.
\end{enumerate}
\emph{Paper Structure.}  
The remainder of this paper is organized as follows. Section~\ref{related} illustrates the background and reviews related work. Section~\ref{method} describes the methodology adopted for our gray literature investigation. Section~\ref{eval} presents the findings that address our research questions. Section~\ref{disc} discusses our results, along with the corresponding implications and threats to validity. 
Finally, Section~\ref{conclusion} concludes the paper and outlines directions for future work.
\section{Background \& Related Work} \label{related}
Fairness in AI/ML systems has been conceptualized in diverse ways, without a single universal definition~\cite{van2020ethical}. Some studies frame fairness as a measurable property, relying on statistical criteria such as group or individual fairness~\cite{chen2024fairness,baresi2023understanding,ferrara2024fairnessinterview,pham2025fairness,ramadan2025towards}, while others emphasize its social and organizational dimensions~\cite{deng2023investigating}, or recommend contextual definitions~\cite{mccormack2024ethical,voria2024catalog,ferrara2024refair,ferrara2024fairness}. As fairness becomes a key aspect of responsible AI, researchers have introduced approaches such as \textit{fairness by design} and \textit{accountability by design}, which integrate fairness into early stages of the SDLC through diverse data collection, auditing, and continuous monitoring~\cite{baresi2023understanding,gunasekara2025systematic}. 
Further, not all fairness definitions can be met simultaneously, as shown by the impossibility theorem~\cite{chouldechova2017fair}, which states that fairness is only satisfied under the assumption that the algorithm is perfect and groups are identical. This, however, is an unrealistic assumption given the unpredictability of input data, and the model behavior itself.
Building on this line of work, we define \textit{fairness requirements} as explicit, context-dependent, and verifiable statements that specify how an AI system should ensure equitable treatment of individuals or groups, prevent discriminatory outcomes, and uphold fairness principles throughout its life cycle~\cite{baresi2023understanding,ferrara2024refair,ramadan2025towards}. 

Although fairness requirements are increasingly formalized, their implementation remains fragmented across organizations. Most industry efforts rely on informal practices such as checklists or team discussions rather than systematic specification and validation~\cite{deng2023investigating}. Frameworks like Fair CRISP-DM~\cite{singh2022fair} advocate fairness integration across all SDLC phases, yet most initiatives still concentrate on de-biasing during model development. As a result, fairness is often treated as an afterthought rather than a foundational specification~\cite{holstein2019improving, baresi2023understanding}. Studies indicate that many cases of algorithmic discrimination stem from the absence of explicit and verifiable fairness requirements and the lack of systematic mechanisms for identifying fairness-related risks in early artifacts~\cite{ramadan2025towards, ferrara2024refair}. Because fairness knowledge is scattered across legal, ethical, and technical domains, engineers face challenges in defining measurable and context-aware conditions~\cite{ramadan2025towards, chen2024fairness}. Therefore, fairness requirements must be explicitly defined, operationalized, and traced throughout the SDLC to ensure that fairness becomes a proactive rather than reactive design property.

Although~\citet{holstein2019improving} found via 35 semi-structured interviews and a survey of 267 ML practitioners that the academic literature focused mainly on algorithmic de-biasing, more recent work has addressed this gap by investigating data-related bias such as unrepresentative, imbalanced, or ethnically skewed training data bias~\cite{van2020ethical,cheng2021socially,baresi2023understanding,joshi2021ai,yang2024survey,sorathiya2024ethical,jakobssonrequirement,holstein2019improving,de2025software,fabris2025fairness,ferrara2024fairnessinterview}, reliance on historical, prejudicial data~\cite{ferrara2024refair, ryan2023integrating,ryanfairness}
, lack of transparency regarding a dataset's origins~\cite{pushkarna2022data}, AI feature bias (e.g., containing potential racial biases)~\cite{deng2023investigating,gunasekara2025systematic}, data formalization, and data misuse~\cite{cheng2021socially}.
Other causes of fairness requirement violation are associated with flaws in algorithms (e.g., choice of the  algorithm, algorithmic design)~\cite{van2020ethical,cheng2021socially, baresi2023understanding,gunasekara2025systematic,singh2022fair,de2025software,fabris2025fairness}, human bias (a.k.a., user bias~\cite{ferrara2024fairness}), systemic inequality ~\cite{van2020ethical,cheng2021socially}, overreliance on AI~\cite{sorathiya2024ethical,singh2022fair}, and organizational challenges~\cite{deng2023investigating,rakova2021responsible} (e.g., ineffective collaboration and lack of resources or awareness)~\cite{cheng2021socially,baresi2023understanding,deng2023investigating,holstein2019improving,de2025software,banks2025multiple}.

These violations lead to social and representational harms, such as discrimination, stereotypes, inconsistent assessments, ethical tensions, financial loss, and reputational damage~\cite{baresi2023understanding, gunasekara2025systematic, sorathiya2024ethical, jakobssonrequirement, holstein2019improving, deng2023investigating, singh2022fair, ferrara2024fairness, ramadan2025towards, rakova2021responsible, ryan2023integrating, ryanfairness, tang2025towards, pant2025navigating}, affecting critical domains such as healthcare~\cite{voria2024catalog, yang2024survey, gunasekara2025systematic,ryanfairness}. For instance, a higher accuracy is shown for light-skinned compared to dark-skinned subjects in diabetic retinopathy~\cite{yang2024survey} and incorrect treatment recommendations are therefore provided to patients, affecting the patient safety and his well-being~\cite{gunasekara2025systematic}. 

While prior work, \newtext{notably a recent survey~\cite{chen2024fairness}, in which technical fairness testing workflow and components are given the most focus}, addresses some of the fairness requirement aspects studied in our paper, none offers a comprehensive view of all the key dimensions explored in this study.
Together, these findings highlight the need to better understand how fairness requirements are defined, managed across the SDLC phases, and violated in practice, motivating our investigation of the gray literature.

\section{Research Method} \label{method}
Our study follows established guidelines for conducting gray literature reviews in software engineering, which emphasize systematic planning, source selection, quality assessment, data extraction, and synthesis~\cite{garousi2020benefitting}. In particular, we followed the methodological structure presented in prior work on fairness debt~\cite{Sotolani2023} and adapted it to focus on fairness requirements in AI and ML systems. Following these guidelines, our process involved four main phases: defining research questions, systematically collecting relevant gray literature, extracting data through a structured review process, and conducting content analysis to synthesize the findings. Each of these phases is described below.

\subsection{Research Questions}\label{RQs}
Fairness requirements lack a universal definition, particularly in AI-based systems. Such requirements are highly context-dependent and may even vary within the same domain~\cite{baresi2023understanding,singh2022fair,holstein2019improving}.
This inconsistency complicates the identification of fairness requirements during AI/ML system development. 
This study explores how fairness requirements are defined in AI/ML, how they are managed during the SDLC, and the primary causes/effects of fairness requirements violations in such systems according to the gray literature. To address this objective, we refined our overarching goal into the following specific research questions:
    \begin{itemize}
    \item \textbf{RQ1:} How are fairness requirements defined in AI/ML within gray literature?
     \item \textbf{RQ2:}  How are fairness requirements managed (e.g., identified, formalized, documented, and addressed) throughout the SDLC of AI/ML systems?
      \item \textbf{RQ3:} What are the primary causes of fairness requirement violations in AI/ML
     systems? 
     \item \textbf{RQ4:} What practical effects or consequences of fairness requirement violations are reported in the gray literature on AI/ML systems?
\end{itemize}

\subsection{Data Collection}

To collect data from the gray literature, we adopted a query-based approach commonly used in recent software engineering reviews~\cite{baltes202040,freire2024eliciting,Sotolani2023}. This method involved launching targeted search queries to systematically retrieve publicly available sources (also referred to as white papers)\footnote{White papers refer to well-structured, non–peer-reviewed documents such as blog posts, reports, and technical articles.} using the Google Custom Search Engine (programmable Google API)~\cite{googleCustomSearch} that discussed fairness in software requirements and specifications.

\subsubsection{Query Design}\label{querydesign}
We first created an initial search query that consists of looking into the overall context of this study (fairness requirements in AI/ML systems): \emph{fairness AND (requirement OR specification) AND (AI OR artificial intelligence OR ML OR machine learning)} (see the first row in Table~\ref{queries}), combining fairness with requirements (or specifications). 
We initially retrieved the top $\initialResults$ results but found that they were not sufficiently relevant to adequately cover our research questions RQ1--RQ4. 
We then refined the initial query, by creating and launching $\searchStrings$ additional search queries (see the last three rows in Table~\ref{queries}) that focus on the remaining key aspects of this study:
the fairness requirements in the SDLC, the primary causes of the violation of such requirements in AI/ML systems, and the corresponding effects/ consequences based on a gray literature investigation. Overall, we considered the top $\topResults$ results from each search query, resulting in a total of $\searchStrings \times \topResults + 100 = \totalResults$ relevant web addresses. All searches were performed
in July 2025, and all metadata and extracted data are available in
our replication package\cite{graylitspreadsheet2025}.

\subsubsection{Inclusion Criteria}\label{inc}
As our study insights are retrieved from the gray literature investigation, we only included publicly available web addresses with i) complete, ii) non-duplicate, iii) English-based and iv) non-academic content. 
Any document that does not meet all the above inclusion criteria (e.g., incomplete or peer-reviewed document such as research paper, book chapter, and reviews) is considered part of the exclusion category and therefore was omitted from further analysis.
Further, we conducted a parallel site-level review during the data collection setup. More in detail,
as part of our API-based retrieval process, our script continuously refined the list of excluded sites based on prior API responses and domain knowledge. 
This iterative filtering helped us avoid retrieving non–gray literature, such as peer-reviewed papers hosted under proceedings or journal sections of university websites.
Subsequently, two researchers independently reviewed all web addresses associated with each final query execution (see Table~\ref{queries}) by first skimming and scanning them, followed by a full reading when the content required deeper investigation or verification.
At the end, each article was classified as either \emph{included} or \emph{excluded}, with reasons for exclusion recorded in the shared Excel sheet~\cite{graylitspreadsheet2025} to ensure transparency and traceability (see Columns `Included / Excluded' and `Reason for Exclusion' under column `Data Collection Phase – Outcome' in the tab associated with each search query). Any disagreements were resolved through consensus meetings, with a third researcher consulted when needed. While our initial selection excluded white papers that did not match all the required search query keywords, we observed that some excluded white papers used minor linguistic variations (e.g., `fair' instead of `fairness') while still addressing the remaining core concepts of the query. As such, we revisited these exclusions and included materials where semantic alignment with the original intent was evident. This step ensured conceptual coverage without losing methodological rigor.
Following consensus, 103 web addresses were selected for data extraction.

\begin{table}[h]
\centering
\scriptsize
\caption{Software fairness requirements search keywords}
\rowcolors{2}{white}{gray!10}
\begin{tabular}{p{1.05cm}|p{.3cm}|p{6cm}}
\hline
\rowcolor{gray!50}
\textbf{Aspect} & \textbf{Q.ID} & \textbf{Search Query} \\
\hline \rowcolor{white}
Fairness \text{Requirement} Definition & 0 & 
\textbf{fairness AND (requirement OR specification) AND (AI OR artificial intelligence OR ML OR machine learning)} \\
\hline \rowcolor{gray!10}
Fairness \text{Requirements} in SDLC & 1 & 
fairness AND (requirement OR specification) AND (AI OR artificial intelligence OR ML OR machine learning) AND 
\textbf{(SDLC OR software development life cycle OR requirement elicitation OR design OR implementation OR development OR testing OR deployment OR maintenance)} \\
\hline \rowcolor{white}
\text{Causes of} Fairness \text{Requirement} Violation & 2 & 
fairness AND (requirement OR specification) AND (AI OR artificial intelligence OR ML OR machine learning) \textbf{AND 
(cause OR reason OR factor OR challenge) AND (unsatisf* OR violat*))} \\
\hline \rowcolor{gray!10}
\text{Effects of} Fairness \text{Requirement} Violation & 3 & 
fairness AND (requirement OR specification) AND (AI OR artificial intelligence OR ML OR machine learning) \textbf{AND 
(effect OR impact OR consequence)} \\
\hline
\end{tabular}
\label{queries}
\end{table}

\subsection{Data Extraction}
Two researchers reviewed the 103 white papers to identify \emph{record units}---short, relevant text segments reflecting aspects of fairness in AI/ML systems aligned with our research questions (see Table~\ref{dataExtraction}). 
Papers were retained only if they addressed at least one aspect, resulting in 38 articles being excluded, leaving 65 articles for analysis. 
A third researcher subsequently reviewed the decisions to confirm consistency and reliability. Disagreements were further discussed and resolved collaboratively to strengthen the validity of the inclusion process.

\begin{table}[h]
\centering
\scriptsize
\caption{Key aspects for inclusion during data extraction}
\rowcolors{2}{white}{gray!10}
\begin{tabular}{p{1.9cm}|p{4.8cm}|p{.6cm}}
\hline
\rowcolor{gray!50}
\textbf{Aspect} & \textbf{Description} & \textbf{Related RQ} \\
\hline \rowcolor{white}
Fairness Definition in AI/ML Requirements \text{context} & How the reference defines or conceptualizes \text{fairness} in AI/ML systems requirements, if applicable & RQ1 \\
\hline \rowcolor{gray!10}
Fairness \text{Requirement} Practices & Mentions of identifying, formalizing, documenting, or incorporating fairness requirements in the SDLC & RQ2 \\ 
\hline \rowcolor{white}
\text{Causes of Fairness} \text{Requirement Violations} & Reports technical, societal, or organizational causes for fairness violation & RQ3 \\ 
\hline \rowcolor{gray!10}
Effects of \text{Fairness} \text{Requirement Violations} & Reports consequences or harms from fairness requirement violations in practice & RQ4 \\
\hline
\end{tabular}
\label{dataExtraction}
\end{table}

The outcome of this step is aligned with the spreadsheet findings, under column `Data Extraction Phase – Outcome' in the tab associated with each search query. Intuitively, white papers that were overly general and lacked substantive discussion of key aspects from our research questions were excluded.

\subsection{Data Analysis}
In total, 686 recorded units were extracted from the 65
articles selected from the data extraction step. Each recorded unit addressed at least one of our research questions RQ1--RQ4 (see Section~\ref{RQs}). More in detail, we collected 78 recorded units addressing RQ1, 202 addressing RQ2, 231 addressing RQ3, and 175 addressing RQ4.
For instance, ``
High-profile missteps such as the Tay chatbot incident, where a Microsoft AI chatbot learned toxic behavior from public interactions on social media''
answers RQ4, demonstrating how fairness requirement violation can erode confidence in AI systems, leading to loss of public trust and legitimacy. We conducted a content analysis~\cite{krippendorff2018content} using manual coding~\cite{seaman1999qualitative} to identify information on the consideration of fairness requirements in SDLC, as well as the causes, and/or effects of the violation of such requirements. More in detail, the content analysis step consists of the three following steps: 

\subsubsection{Initial Coding (Pilot)} Two researchers independently coded two randomly selected sources (14 recorded units) to test the coding protocol, where each \emph{recorded unit} addresses at least one research question.
For instance, the recorded unit:
``Maintaining automatically generated AI system logs, to the extent such logs are under their control, for a specified period'' is associated with RQ2, and coded as
\emph{Fairness Monitoring \& Evaluation}.
Similarly, the recorded units ``Bias, often based on race, gender, age or location, has been a long-standing risk in training AI models'' and ``...show that these generative AI models manifest biases that may be inherent in the training data collected, reflecting societal stereotypes'', are both associated with RQ3, and coded as
\emph{Input / Data \& Representation Bias}.
Disagreements were discussed, and a shared understanding of category definitions was established.
For example, one researcher identified the recorded unit ``Social scoring systems—systems that evaluate or classify individuals based on their social behavior'' as both a definition of fairness requirement and a cause of fairness requirement violation, answering both RQ1 and RQ3, respectively. After a consensus meeting, both researchers agreed that the recorded unit indeed fits under both categories; while one researcher saw it as indirectly defining fairness by prohibiting bias based on social behavior, the other researcher noted it also implies causes of fairness violations since relying on social behavior to classify individuals can produce bias.

\subsubsection{Calibration Round}
The annotators conducted a calibration
round with 20 articles (83 recorded units). During this phase, both annotators coded the same subset of data, compared results, and discussed discrepancies to refine code definitions and strengthen inter-coder reliability before proceeding to the full analysis. 

\subsubsection{Full Coding}
In the final iteration, one annotator coded the remaining 589 (686 - (14 + 83)) recorded
units, and a second annotator reviewed the coding.
At this stage, consensus meetings were also held in case of disagreement to finalize the set of codes generated from all units recorded in this study. 
For example, the unit 
``Equality of Opportunity is defined as a fairness criterion that requires a classifier to have equal true positive rates across different protected groups'' associated with RQ1 was initially interpreted differently by the co-authors: one coded it as Formal / Causal and Model-based Fairness, emphasizing theoretical fairness frameworks, while the other coded it as Predictive Fairness / Statistical Parity / Equalized Probability, focusing on measurable statistical criteria. After discussion, the co-authors ultimately agreed on the latter code as the most accurate one, as it best captures the operational, data-driven aspect of the fairness criterion described in the quote.

To explore the relationships between the definitions of the fairness requirement (RQ1), the fairness requirement management practices in the SDLC (RQ2), the causes of violations of the fairness requirements (RQ3), and the corresponding violation effects (RQ4), we analyzed the co-occurrence within the same recorded unit for the 43 units that address more than one research question in our gray literature (see sheet `Co-occurrence Analysis'~\cite{graylitspreadsheet2025}).
Each recorded unit was reviewed to determine the code(s) under which it falls in addressing each research question. For example, a quote might define a fairness requirement (RQ1), along with the cause(s) of its violation (RQ3), and/or the corresponding violation effect (RQ4).
For instance, the quote ``To address this, AI...supports decision making by highlighting overlooked features or potential flaws in judgment...it also carries risks, such as overreliance on AI, which could lead to deskilling and a significant loss of medical expertise'' answers both RQ3 and RQ4, by highlighting \emph{human \& judgment factors}, such as overreliance on AI as cause of fairness requirement violation, and \emph{professional \& societal harms}, like loss of medical expertise as consequence of the fairness requirement violation.
These co-occurrence relationships were recorded in a mapping table and aggregated across all quotes (see `Co-occurrence analysis' sheet in the shared material~\cite{graylitspreadsheet2025}. 
Not limited to specific combinations of RQs, the same logic was applied to generate mappings between fairness requirement definition / management in the SDLC 
(RQ1--RQ2) 
and fairness requirement violation causes and effects (RQ3--RQ4), with co-occurrence frequencies illustrating the strongest relationships.

To assess inter-rater reliability, we computed Cohen’s kappa coefficient
(k)~\cite{mchugh2012interrater}, which ranges from -1 (complete
disagreement) to +1 (perfect agreement). We used the ReCal2
tool~\cite{freelon2010recal} to compute $K$ after the three data analysis iterations. For definition of fairness requirements (RQ1), we obtained $K= 0.76$ and for both fairness requirement management in SDLC (RQ2), and root causes of fairness requirement violation (RQ3), we obtained $k=0.73$, showing
good agreement between the coders.
For the consequences / effects of fairness requirement violation (RQ4), the coefficient was $K= 0.84$, reflecting strong agreement between the coders.
\section{Results} \label{eval}
\begin{figure}[h]
  \centering
\includegraphics[width=0.4\textwidth,height=0.15\textheight]
{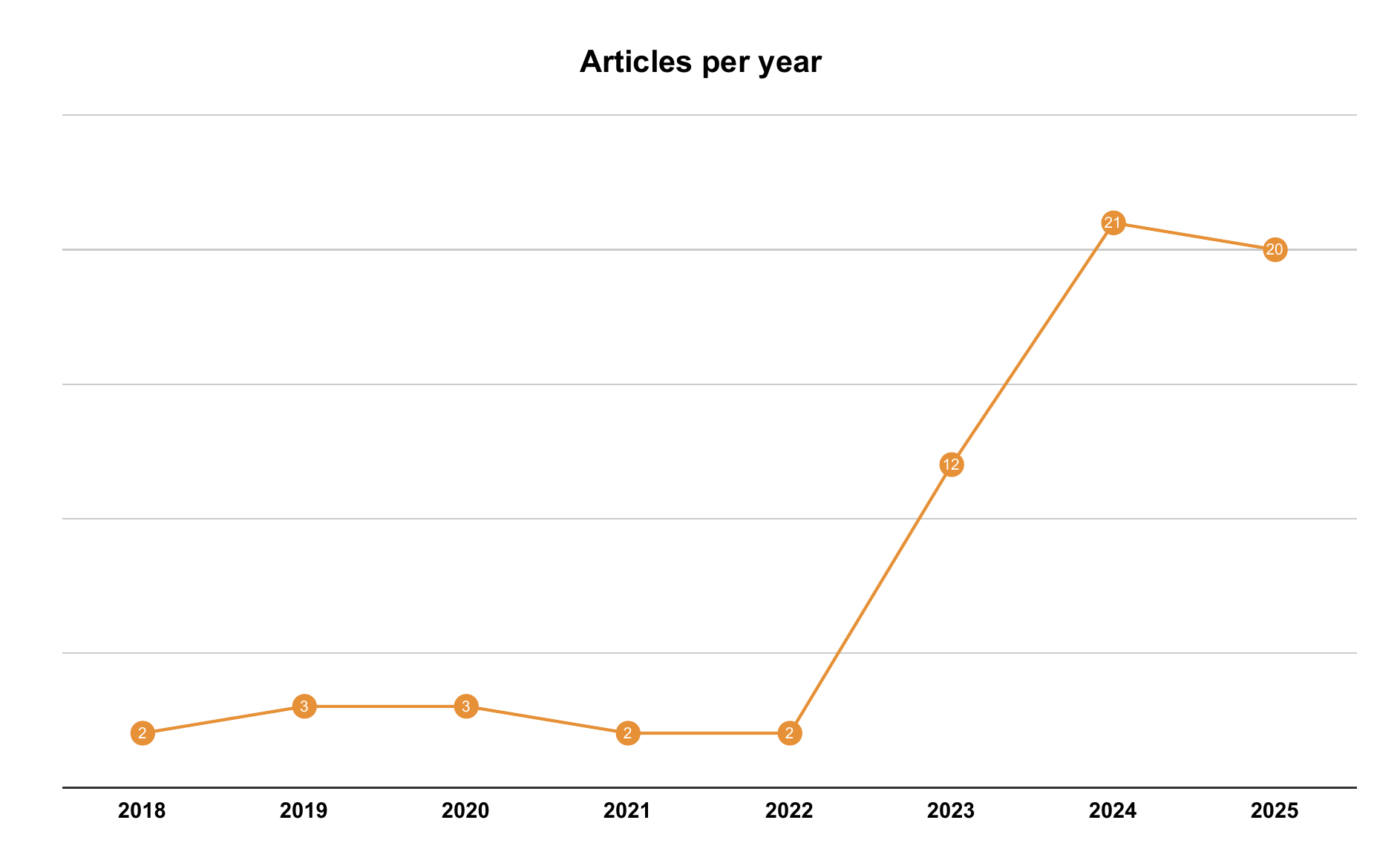}
  \caption{Number of articles per year}
  \label{referencesperyear}
\end{figure}

We analyzed 65 articles (686 recorded units) on fairness requirements. Figure~\ref{referencesperyear} shows their yearly distribution, revealing a
rising trend that peaks in 2024. The following sections
present our results, organized by research question.  
For space limitation, we report the top five findings from the gray literature for each of the four research questions (RQ1--RQ4). The full findings are reported in the `Final\_Results' sheet of our replication package~\cite{graylitspreadsheet2025}.

\subsection{RQ1: How are fairness requirements defined in AI/ML
within gray literature?}
In Table~\ref{rq1table}, we report the top five fairness requirement definitions commonly adopted in the gray literature investigation, along with the number of recorded units (Column \emph{\# Rec. Units}) and the references that support each definition (Column \emph{References}). The most common definition collected from 18 recorded units and cited by 13 references frames fairness as non-discrimination or equal treatment, emphasizing ethical and normative principles. Other commonly cited definitions include contextual, ethical, and normative fairness (18 recorded units; 11 references), mainly stating that fairness in AI is context-dependent (e.g., Healthcare, social scoring systems), showing that AI fairness requirements extend beyond non-discrimination, reaching into philosophical, cultural, historical, political, and ethical contexts, where the fairness requirement definition varies accordingly.
AI Fairness requirement definition is also relative to sensitive attributes (13 recorded units; 10 references), drawing attention to specific sensitive attributes, notably gender, ethnicity, race, age, and disability. Fairness requirements are less frequently defined as formal, causal, and model-based fairness (9 units; 8 references),
where, for for instance, 
causal fairness analyzes the indirect impact of sensitive attributes on AI system decisions.
Further, fairness requirements in AI are also defined as predictive fairness / statistical parity / equalized probability (8 units; 5 references), 
emphasizing fairness through statistical criteria such as the probability of positive outcomes across groups and equal True Positive Rates (TPR) across different application domains such as healthcare, where patients need to be equally treated under the same health characteristics regardless of their disability, gender or any other sensitive attribute. 
\begin{table}[!h]
\centering
\scriptsize
\caption{Top five fairness requirement definitions}
\begin{tabular}{p{2cm}|p{.45cm}|p{4.9cm}}
\hline
\rowcolor{lightgray} 
\textbf{Code} & \textbf{\text{\# Rec.} Units} & \textbf{References} \\
\hline
\text{Fairness as Non-} \text{Discrimination / Equal} Treatment & \centering 18 & GL027, GL050, GL077, GL086, GL091, GL098, GL106, GL122, GL125, GL128, GL199, GL271, GL277 \\
\hline  \rowcolor{gray!10}
Contextual, Ethical, and Normative Fairness & \centering 18  & GL027, GL083, GL098, GL104, GL105, GL106, GL125, GL187, GL277, GL348, GL392 \\ \hline

Fairness Relative to \text{Sensitive Attributes} & \centering 13 & GL027, GL031, GL038, GL073, GL105, GL187, GL199, GL202, GL277, GL359 \\ \hline  \rowcolor{gray!10}
\text{Formal / Causal} \text{and Model-Based} Fairness
& \centering 9 & GL047, GL066, GL098,GL125, GL187, GL321, GL348, GL356\\ \hline
\text{Predictive Fairness /} \text{Statistical Parity /} Equalized probability & \centering 8 & GL066, GL073, GL125, GL348, GL392\\ \hline
\end{tabular}
\label{rq1table}
\end{table}

The answer to RQ1 shows that fairness requirements in AI are multifaceted and interpreted in various ways, covering ethical, non-discriminative principles, context-based interpretations spanning various application domains, sensitive attribute-based considerations (e.g., gender, race, age, disability, etc), and technical and formal model-based definitions. 
\newtext{While non-discrimination definitions dominate, technical and formal model-based ones are less frequent, highlighting a gap between conceptual discussion and operational definitions. Further, fairness requirements are interpreted across domains, showing context-based adaptation, with operational definitions less common and varying across domains.}

\subsection{RQ2: How are fairness requirements managed throughout the SDLC of AI/ML systems?}
Based on Table~\ref{rq2table}, 
the gray literature references most frequently address fairness in the model training phase and associated bias mitigation techniques (55 recorded units; 23 references), which focus on methods such as pre-processing (e.g., re-sampling, re-weighting, and de-biasing during model training to reduce prediction bias), in-processing (e.g., incorporation of fairness constraints during model training), post-processing, adversarial debiasing, and model adaptation (i.e., modification of a trained model to improve its fairness) during the software development stage of the SDLC. Another common approach is fairness monitoring and evaluation (31 recorded units; 22 references), ensuring a continuous assessment of models to detect and address fairness issues throughout the deployment and maintenance phases. Fairness through data collection and preparation practices (30 recorded units; 19 references), including the collection of diverse and representative data, the removal of identifiable discriminatory bias in data, Combinatorial Interaction Testing (CIT), and securing data so that it remains protected against manipulation, given that untrustworthy data can introduce hidden biases that undermine fair outcomes.  
Less frequently, fairness requirements are expressed through requirements and policy principles (14 recorded units; 10 references), where fairness is guided by 
ethical frameworks, 
AI requirements and policies ensuring non-discrimination, equal treatment, and respect for human dignity through explicit guidelines and documented assurance processes.
Finally, fairness testing and evaluation (9 recorded units; 7 references) assesses 
AI and software systems for bias and ethical implications, using proper fairness metrics (e.g., demographic parity,  counterfactual fairness, accuracy parity, etc), analytic tools (e.g., LANGTEST), post-hoc analysis tools, and testing on different, real-world datasets to evaluate fairness across diverse groups and contexts. 
\begin{table}[h]
\centering
\scriptsize
\caption{Top five fairness requirement management tasks w.r.t. the SDLC stages}
\begin{tabular}{p{2cm}|p{.45cm}|p{4.9cm}}
\hline
\rowcolor{lightgray} 
\textbf{Code} & \textbf{\text{\# Rec.} Units} & \textbf{References} \\
\hline
\text{Fairness in Model} Training \& Bias Mitigation Techniques & \centering 55 & GL031, GL059, GL066, GL067, GL073, GL077, GL098, GL105, Gl106, GL120, GL128, GL132, GL135, GL187, GL214, GL215, GL241, GL300, GL321, GL348, GL356, GL359, GL392 \\ \hline  \rowcolor{gray!10}
Fairness Monitoring \& Evaluation & \centering 31 &  GL010, GL017, GL020, GL048, GL073, GL077, GL083, GL086, GL097, GL105, GL106, GL120, GL128, GL135, GL141, GL160, GL180, GL187, GL202, GL271, GL335, GL346 \\ \hline 
Fairness Through Data Practices & \centering 30 & GL017, GL027, GL031, GL047, GL059, GL077, GL083, GL098, GL121, GL125, GL128, GL141, GL172, GL180, GL187, GL300, GL346, GL348, GL392 \\ \hline  \rowcolor{gray!10}
Requirements \& Policy: Fairness Principles & \centering 14 & GL017, GL027, GL086, GL104, GL125, GL132, GL160, GL236, GL241, GL277 \\ \hline
\text{Fairness Testing \&} Evaluation  & \centering 9 & GL059, GL077, GL106, GL141, GL160, GL249, GL388 \\ \hline 
\end{tabular}
\label{rq2table}
\end{table}

The answer to RQ2 shows that fairness is often treated as a technical issue, with bias mitigation and monitoring being mostly prioritized, while governance and early-stage practices receive less attention.
This pattern reveals gaps in proactive measures and indicates that current efforts are primarily reactive, highlighting areas where upstream interventions could better support fairness in practice.
\newtext{RQ2 findings point to a reactive orientation, where fairness attention is primarily concentrated during or after AI model development, leaving early-stage design and governance with a least priority or less focus. RQ2 therefore suggests a proactive orientation for better fairness adoption in AI software, starting from early SDLC stages.}

\subsection{RQ3: What are the primary causes of fairness requirement violations in AI/ML systems?}
As shown in Table~\ref{rq3table}, the most frequent cause of fairness requirement violation is \emph{Input / data \& representation bias} (112 units; 36 references), where imbalanced or unrepresentative datasets lead to discriminatory outcomes. Examples include arrest records that overrepresent certain communities, employment data dominated by one gender, and facial recognition datasets containing mostly white individuals. Related data issues such as labeling and annotation errors, historical and proxy biases, variable omission, and self-selection further reinforce unfairness. A second major cause is \emph{algorithmic or model design bias} (48 units; 26 references), which results from design choices or model architectures that unintentionally embed unfairness. These include causal shortcuts, opaque or non-interpretable models, overemphasis on accuracy metrics, and trade-offs between fairness, accuracy, and privacy. \emph{Human \& judgment factors} (28 units; 15 references) also contribute, particularly through subjective labeling, biased feature selection, overreliance on AI, and lack of diversity within development teams. Less frequently, \emph{evaluation \& transparency gaps} (8 units; 6 references) limit the ability to detect or mitigate unfairness due to limited data sharing, inadequate testing, and lack of standard benchmarks. 
Finally, \emph{societal / structural biases} (7 units; 6 references) occur when AI systems reproduce existing inequalities in certain domains such as predictive policing, hiring, and credit scoring, where models trained on biased historical data perpetuate systemic discrimination. 

\begin{table}[h]
\centering
\scriptsize
\caption{Top five causes of fairness requirement violation}
\begin{tabular}{p{1.8cm}|p{.45cm}|p{5cm}}
\hline
\rowcolor{lightgray}
\textbf{Code} & \textbf{\text{\# Rec.} Units} & \textbf{References} \\
\hline
\text{Input / Data \&} \text{Representation Bias} & \centering 112 & GL010, GL027, GL031, GL037, GL040, GL047, GL050, GL059, GL086, GL097, GL098, GL106, GL121, GL125, GL128, GL135, GL141, GL153, GL172, GL180, GL187, GL215, GL236, GL241, GL249, GL271, GL277, GL278, GL316, GL335, GL348, GL356, GL359, GL363, GL388, GL392\\ \hline  \rowcolor{gray!10}
\text{Algorithmic / Model} Design Bias & \centering 48  & GL010, GL027, GL047, GL050, GL067, GL077, GL0086, GL097, GL098, GL105, GL125, GL128, GL135, GL160, 
GL192, GL215, GL236, GL241, GL273, GL321, GL335, GL348, GL356, GL359, GL363, GL392 \\ \hline
\text{Human \& Judgment} Factors & \centering 28 & GL027, GL086, GL098, GL135, GL141, GL160, GL202, GL207, GL251, GL271, GL335, GL348, GL356, GL359, GL388\\ \hline  \rowcolor{gray!10}
Evaluation \text{\& Transparency Gaps} & \centering 8 & GL027, GL141, GL153, GL180, GL321, GL335\\ \hline
\text{Societal / Structural} Bias & \centering 7 &
GL050, GL066, GL083, GL132, GL207, GL316\\ \hline 
\end{tabular}
\label{rq3table}
\end{table}

The answer to RQ3 shows that fairness requirement violations in AI and ML systems arise from interconnected technical, human, and societal causes. Addressing these issues requires improving data quality, model design, and evaluation practices, as well as promoting diverse teams and greater awareness of the social contexts in which AI operates.
\newtext{RQ3 findings show that the gray literature focuses on technical causes of fairness violations, while non-technical causes receive less attention, potentially leaving a gap in effectively achieving AI model fairness from a practical standpoint.}

\subsection{RQ4: What practical effects or consequences of fairness requirement violations are reported in the gray literature on AI/ML systems?}

As per results in Table~\ref{rq4table}, the most frequently discussed consequence of fairness requirement violations is harm from biased predictions (111 units; 35 references), where unfair model outputs lead to worse outcomes for underrepresented groups, who may receive less accurate care compared to the majority populations. 
For instance, in healthcare, women and minorities may face poorer predictive accuracy, and young female doctors may be underrepresented in training data, making models less reliable for them. Similarly, non-native speakers and people with speech disabilities can experience accuracy disparities in large pre-trained models. 
Another significant effect of fairness requirement violation is professional and societal harms (39 units; 22 references), which captures broader consequences such as workplace inequities, reduced opportunities, or social exclusion that result from biased system decisions. 
For example, overreliance on AI in healthcare can lead to deskilling and loss of medical expertise. Further, biased systems like the i) Apple Card, where the  algorithm determines higher credit card limits to males than females, even for couples who share their finances, ii)
META’s AI-based advertisemenet, which shows bias towards race, gender, and religion, and iii) 
Amazon’s hiring tool, which favorites men over women. 
Additionally,
AI systems can amplify prejudices because their recommendation algorithms predict which content will get more clicks and views, boosting harmful or racist posts, as seen when an Indigenous Australian football player faced harassment and ended quitting both football and social media.
Stereotype and role reinforcement (16 units; 11 references) highlights the risk of AI systems perpetuating or amplifying harmful social stereotypes.
For example, men are generally associated men with leadership roles (e.g., CEOs, engineers), individuals with small eyes as predicted to be Asians, and nurses are predicted to be women by default, according to generative models decisions.  
Less frequently, violations are linked to data and privacy risks (11 units; 8 references), where fairness concerns intersect with misuse of sensitive data or breaches of confidentiality. 
For example, unfair AI decisions can lead to unequal privacy risks especially for unprivileged, discriminated groups, as trying to investigate why the AI/ML model has been unfair can expose sensitive information from training data associated with such groups. In this case, AI decisions are very likely to cause privacy violations.
Finally, loss of trust and legitimacy (6 units; 5 references) reflects the erosion of stakeholder confidence when AI systems fail to meet fairness expectations.
For example, when Microsoft AI Chatbot learned toxic behavior from social media, public trust in AI has significantly decreased. Further, opaque or unfair decisions of AI systems from healthcare and smart home applications led to distrusting such technologies.
\begin{table}[h]
\centering
\caption{Top five effects of fairness requirement violation}
\scriptsize
\begin{tabular}{p{1.2cm}|p{.45cm}|p{5.8cm}}
\hline
\rowcolor{lightgray} 
\textbf{Code} & \textbf{\text{\# Rec.} Units} & \textbf{References} \\
\hline
Harm from Biased \text{Predictions} & \centering 111  & GL027, GL031, GL037, GL040, GL066, GL067, GL083, GL088, GL098, GL104, GL105, GL121, GL122, GL125, GL128, GL135, GL141, GL153, GL180, GL187, GL202, GL207, GL214, GL236, GL249, GL277, GL316, GL321, GL335, GL346, GL348, GL356, GL359, GL363, GL388\\ \hline \rowcolor{gray!10}
Professional \& Societal Harms & \centering 39 & GL027, GL066, GL067, GL077, GL086, GL104, GL132, GL160, GL172, GL180, GL199, GL207, GL236, GL241, GL249, GL251, GL277, GL278, GL316, GL335, GL348, GL359\\ \hline
Stereotype / Role Reinforcement & \centering 16 & GL040, GL067, GL128, GL236, GL241, GL249, GL316, GL348, GL359, GL363, GL388\\ \hline \rowcolor{gray!10}
Data and Privacy Risk & \centering 11 & GL027, GL040, GL105, GL122, GL141, GL192, GL199, GL321\\ \hline
\text{Loss of Trust} \text{and Legitimacy} & \centering 6 & GL017, GL027, GL104, GL135, GL316\\ \hline
\end{tabular} \label{rq4table}
\end{table}

The answer to RQ4 shows that fairness issues in AI/ML systems can go beyond technical errors, affecting people’s opportunities, social perception, and trust in these systems, especially when the corresponding model decisions are impacted by stereotypes, affecting various demographic groups with disadvantaged races, genders, and other marginalized groups.
\newtext{Overall, all consequences of fairness requirement violations, including less frequent moral, reputational or trust related ones, should be carefully considered to better prioritize fairness and prevent discrimination in AI/ML systems, minimizing harms from violating fairness requirements.}
\subsection{Co-Occurrence Analysis}
\begin{figure}[h!]
    \centering
\includegraphics[width=0.35\textwidth]{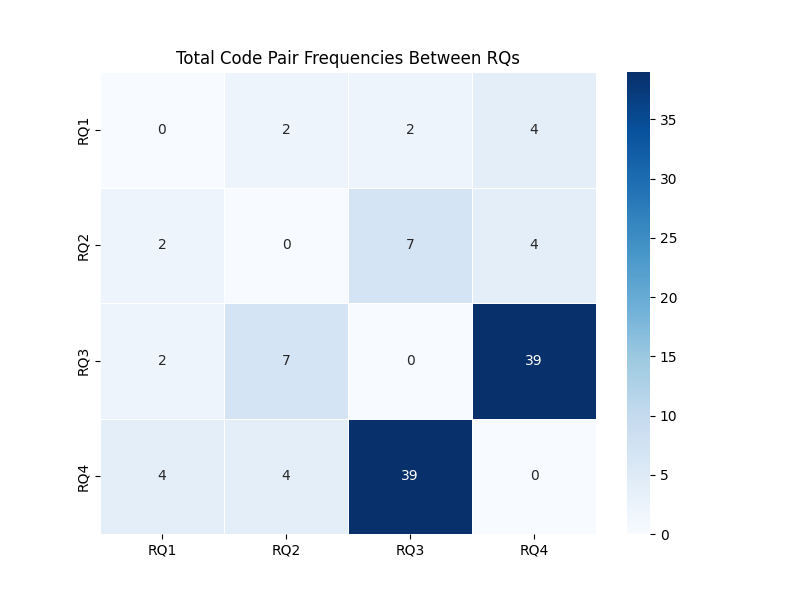}
    \caption{Top code pair co-occurrence matrix (frequency >2)}
    \label{matrix}
\end{figure}

Based on the co-occurrence matrix in figure~\ref{matrix}, the pair of research questions RQ3--RQ4 showed the strongest relationship, followed by RQ2--RQ3, and RQ1--RQ4 / RQ2--RQ4. The heatmap visually highlights this contrast, with darker shading for RQ3--RQ4 indicating frequent co-occurrence between causes and effects of fairness requirement violation, while lighter cells such as those associated with the pairs RQ1--RQ2 point to weaker connections, where, for instance, fairness requirements are defined without considering how they will be managed during the SDLC process.
Building on these patterns, we further investigated the associated codes and their frequencies to uncover underlying thematic links.

\begin{figure}[h!]
    \centering
\includegraphics[width=0.45\textwidth]{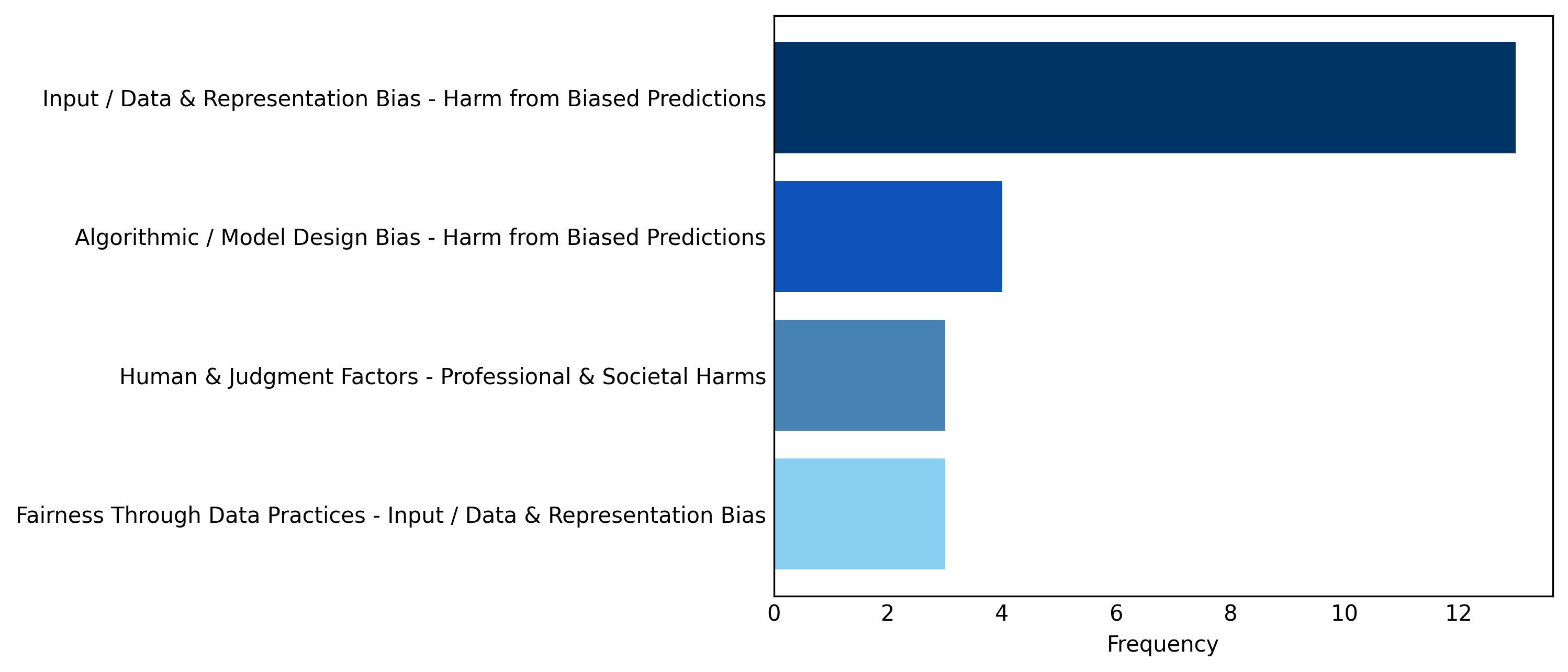} 
    \caption{Top code pair co-occurrences (frequency >2)}
    \label{co-occurrence}
\end{figure}

As depicted in \figurename~\ref{co-occurrence}, most of the frequency counts of code pairs obtained from our study come from the RQ3--RQ4 code pair analysis. 
Among these pairs, we found that \emph{Input / Data \& Representation Bias} and \emph{Algorithmic / Model Design Bias} co-occurred most frequently, both leading to \emph{Harm from Biased Predictions}.
For example, the recorded unit ``Unfairness can be imparted in models because of bias present in training data. Various types of bias, such as annotation bias, historical bias, prejudice bias, etc may lead to unfair models and selective bias towards a particular group'' answering both RQ3 and RQ4, illustrates how \emph{Input/Data \& Representation Bias} directly contributes to \emph{Harm from Biased Predictions}, highlighting the causal link between biased data and the unequal impact of model outcomes on different groups.
Further, \emph{Human \& Judgment Factors} paired less often and were associated with \emph{Professional \& Societal Harms}.
For example, the recorded unit ``To address this, AI...supports decision making by highlighting overlooked features or potential flaws in judgment...it also carries risks, such as overreliance on AI, which could lead to deskilling and a significant loss of medical expertise'' illustrates how \emph{Human \& Judgment Factors} such as clinicians’ reliance on AI can lead to \emph{Professional \& Societal Harms} in the medical domain, leading to loss of expertise, which represents a professional harm.

\section{Discussion}\label{disc}
Despite the various definitions of fairness requirements in AI context and the different insights from studies on violation causes and effects on AI decisions, and their integration in the SDLC, we observed that shared AI fairness contexts originate from different references, with each reference typically covering only one or a few contexts. 
We also did not find studies that have explored all facets of AI fairness across all aspects covered by our research questions (see Section~\ref{RQs}). Our review of the gray literature therefore revealed much broader findings covering a wide range of application areas in multiple real-world contexts, notably healthcare, hiring, criminal justice, and social media (see Table~\ref{contexts} for a full list of domains from the gray literature and the corresponding number of quotes collected per context from Column `\#Quo').
\begin{table}[ht]
\centering
\scriptsize
\caption{AI fairness application contexts and quotes}
\begin{tabular}{p{.99cm}|p{.4cm}|p{5.95cm}}
\hline \rowcolor{gray!50}
\textbf{Context} & \textbf{\#Quo.} & \textbf{Examples} \\
\hline
Healthcare & 50 & Cardiovascular disease (CVD) prediction, medical imaging, Electronic Health Records (EHRs), patient experience \\ \rowcolor{gray!10}
Criminal Justice & 36 & Predictive policing, recidivism prediction, facial recognition \\
Hiring & 37 & Resume screening, job recommendation systems, job postings \\ \rowcolor{gray!10}
Finance & 18 & Credit scoring, face-to-face and algorithmic lending \\
Education & 7 & School performance prediction, algorithmic grading \\ \rowcolor{gray!10}
Social \text{Media} & 35 & Content moderation, user profiling, recommendation algorithms, search engines \\
Smart Homes & 14 & Residential smart home systems \\
\hline
\end{tabular}
\label{contexts}
\end{table}
Building on these observations, our findings support several trends identified in prior research while also extending them in scope and practical relevance. Consistent with the academic literature~\cite{baresi2023understanding,gunasekara2025systematic,deng2023investigating,chen2024fairness,ferrara2024fairness,lu2022towards}, the gray literature frequently frames fairness requirements around non-discrimination, equal treatment, and context sensitivity. However, unlike formal studies that conceptualize fairness primarily through taxonomies~\cite{baresi2023understanding} or technical metrics such as group or individual fairness~\cite{chen2024fairness,ferrara2024fairness,lu2022towards}, or recommended to adopt contextual fairness definitions and strategies~\cite{mccormack2024ethical,voria2024catalog,ferrara2024refair,ferrara2024fairness}, our 
gray literature 
further encompasses diverse fairness definitions across domains, highlighting our novel review-based finding that practitioners interpret fairness as context-dependent, tailored to the needs of specific applications rather than a universal metric.

Further, our results align with prior calls for fairness by design and accountability by design~\cite{baresi2023understanding,gunasekara2025systematic}, while also revealing a continued focus on reactive practices, particularly bias mitigation and model evaluation, instead of proactive fairness specification and governance~\cite{holstein2019improving}. 
The co-occurrence analysis further exposes patterns rarely detailed in academic work, linking biased data and model design flaws directly to real-world harms such as professional exclusion, stereotype reinforcement, and loss of trust. These connections, together with the broader coverage of fairness across multiple domains, offer a novel, practice-grounded perspective on how fairness requirements are defined, managed, and violated throughout the AI/ML software lifecycle.

\subsection{Implications}
\textbf{Research Implications.}
The results highlight the need to bridge conceptual and applied understanding of fairness requirements in AI. While academic studies have formalized fairness as statistical, causal, or design-based constructs, the gray literature demonstrates how practitioners reinterpret these notions as context-dependent and operational requirements. Future research should therefore investigate how fairness requirements evolve from abstract principles into actionable specifications across different SDLC stages. 
Empirical studies can examine how early-stage elicitation, traceability, and validation of fairness requirements affect downstream bias mitigation and auditing. Cross-disciplinary work combining software engineering, human–computer interaction, and social science methods could also clarify how fairness principles are negotiated among stakeholders and embedded in engineering practices.

\noindent \textbf{Practice Implications.}In light of RQ1 findings, we strongly suggest that practitioners should clearly define fairness in their specific context, internally communicate fairness concerns, notably from a sensitive attribute perspective, and establish a shared understanding within teams early in the SDLC to ensure consistent ethical practices across the SDLC.
RQ2 suggests that practitioners should integrate fairness throughout the SDLC, from data collection and preprocessing to model training, testing, and deployment. Further, the combination of technical measures such as bias mitigation techniques and monitoring with clear policies and/or guidelines can help ensure fairness is consistently maintained rather than addressed only after fairness issues arise. 
The causes of fairness requirement violation in RQ3 suggest that practitioners combine technical solutions, such as improved data quality and model design, with processes that reduce human bias and consider broader societal context. Clear internal communication, ongoing monitoring, and transparent reporting help teams address fairness issues effectively throughout development and deployment.
Given the broad effects of the violation of fairness requirements on AI decisions, RQ4 suggests that practitioners should integrate fairness checks into decision making and communication. For instance, promoting transparency and stakeholder engagement can help mitigate negative social and organizational effects on AI decisions.

\subsection{Threats to Validity}
We addressed potential threats to validity by ensuring methodological rigor and transparency throughout all phases of the review. To address \textit{internal validity} threats, we support \textit{credibility} by having two researchers independently conduct data collection, extraction, and coding, with disagreements resolved through consensus meetings and a review by a third researcher. Cross-checking of quotes and assigned codes reduced the risk of extraction or interpretation errors. Further, to mitigate \textit{search bias}, we used the Google Custom Search API with multiple query formulations (see Table~\ref{queries}) and included a wide range of gray literature sources such as blogs, reports, and preprints. 
To strengthen \textit{construct validity}, we complemented fairness-related queries with semantically related terms such as \textit{bias} and \textit{discrimination} to capture relevant materials that used different terminology. \textit{External validity} may be limited since gray literature often reflects specific organizational or regional contexts; however, we used unrestricted global searches and transparent inclusion criteria (see Section~\ref{inc}) to broaden representativeness. Finally, detailed documentation of all procedures and data (see replication package~\cite{graylitspreadsheet2025}) ensures \textit{dependability} and supports reproducibility.
\section{Conclusion and Future Work} \label{conclusion}
In this work, we conducted a gray literature review, in which we show how fairness in AI models is interpreted from ethical principles to technical, model-based definitions, emphasizing non-discrimination and attention to sensitive attributes. We also studied how fairness requirements are managed in the SDLC, finding that this is primarily done through the use of representative data practices, bias mitigation techniques during model training, and continuous monitoring, while early-stage governance and requirements receive less attention.
Key causes of AI fairness violations include biased data, model design choices, human judgment, and societal factors. These violations can lead to biased model predictions, social inequities, stereotype reinforcement, privacy risks, and loss of trust. Overall, addressing fairness in AI systems requires both technical solutions and consideration of human and societal influences. As a future direction, we plan to develop a framework to guide fairness requirements systematically throughout the SDLC.

\bibliographystyle{ACM-Reference-Format}
\bibliography{bibliography}

@inproceedings{Sotolani2023,
  author = {Rodrigo Sotolani and Ronnie de Souza Santos and Felipe Fronchetti and Savio Freire and Rodrigo Spinola},
  title = {Exploring Software Fairness Debt in Gray Literature},
  year = {2023},
}

@article{chen2024fairness,
  title={Fairness testing: A comprehensive survey and analysis of trends},
  author={Chen, Zhenpeng and Zhang, Jie M and Hort, Max and Harman, Mark and Sarro, Federica},
  journal={ACM Transactions on Software Engineering and Methodology},
  volume={33},
  number={5},
  pages={1--59},
  year={2024},
  publisher={ACM New York, NY}
}

@article{baltes202040,
  title={Is 40 the new 60? How popular media portrays the employability of older software developers},
  author={Baltes, Sebastian and Park, George and Serebrenik, Alexander},
  journal={IEEE Software},
  volume={37},
  number={6},
  pages={26--31},
  year={2020},
  publisher={IEEE}
}

@inproceedings{freire2024eliciting,
  title={Eliciting Public Discourse of SE Tool Providers in a Study on Requirements Process Debt-A Different Shade of Gray},
  author={Freire, S{\'a}vio and Maciel, Rita SP and Mendon{\c{c}}a, Manoel and Leite, Julio Cesar},
  booktitle={Proceedings of the XXIII Brazilian Symposium on Software Quality},
  pages={189--198},
  year={2024}
}

@misc{googleCustomSearch,
  author       = {{Google Developers}},
  title        = {{Custom Search JSON API}},
  year         = {2024},
  howpublished = {\url{https://developers.google.com/custom-search}},
  note         = {Accessed: 2025-07-04}
}

@inproceedings{baresi2023understanding,
  title={Understanding fairness requirements for ml-based software},
  author={Baresi, Luciano and Criscuolo, Chiara and Ghezzi, Carlo},
  booktitle={2023 IEEE 31st International Requirements Engineering Conference (RE)},
  pages={341--346},
  year={2023},
  organization={IEEE}
}

@article{chouldechova2017fair,
  title={Fair prediction with disparate impact: A study of bias in recidivism prediction instruments},
  author={Chouldechova, Alexandra},
  journal={Big data},
  volume={5},
  number={2},
  pages={153--163},
  year={2017},
  publisher={Mary Ann Liebert, Inc. 140 Huguenot Street, 3rd Floor New Rochelle, NY 10801 USA}
}

@article{bellamy2019ai,
  title={AI Fairness 360: An extensible toolkit for detecting and mitigating algorithmic bias},
  author={Bellamy, Rachel KE and Dey, Kuntal and Hind, Michael and Hoffman, Samuel C and Houde, Stephanie and Kannan, Kalapriya and Lohia, Pranay and Martino, Jacquelyn and Mehta, Sameep and Mojsilovi{\'c}, Aleksandra and others},
  journal={IBM Journal of Research and Development},
  volume={63},
  number={4/5},
  pages={4--1},
  year={2019},
  publisher={IBM}
}

@article{bird2020fairlearn,
  title={Fairlearn: A toolkit for assessing and improving fairness in AI},
  author={Bird, Sarah and Dud{\'\i}k, Miro and Edgar, Richard and Horn, Brandon and Lutz, Roman and Milan, Vanessa and Sameki, Mehrnoosh and Wallach, Hanna and Walker, Kathleen},
  journal={Microsoft, Tech. Rep. MSR-TR-2020-32},
  year={2020}
}

@article{saleiro2018aequitas,
  title={Aequitas: A bias and fairness audit toolkit},
  author={Saleiro, Pedro and Kuester, Benedict and Hinkson, Loren and London, Jesse and Stevens, Abby and Anisfeld, Ari and Rodolfa, Kit T and Ghani, Rayid},
  journal={arXiv preprint arXiv:1811.05577},
  year={2018}
}

@article{mehrabi2021survey,
  title={A survey on bias and fairness in machine learning},
  author={Mehrabi, Ninareh and Morstatter, Fred and Saxena, Nripsuta and Lerman, Kristina and Galstyan, Aram},
  journal={ACM computing surveys (CSUR)},
  volume={54},
  number={6},
  pages={1--35},
  year={2021},
  publisher={ACM New York, NY, USA}
}

@article{simpson2024parity,
  title={Parity benchmark for measuring bias in LLMs},
  author={Simpson, Shmona and Nukpezah, Jonathan and Brooks, Kie and Pandya, Raaghav},
  journal={AI and Ethics},
  pages={1--15},
  year={2024},
  publisher={Springer}
}

@misc{graylitspreadsheet2025,
  author       = {},
  title        = {Gray Literature Source Review Spreadsheet},
  year         = {2025},
  url          = {https://docs.google.com/spreadsheets/d/1pzIWOHGdAAZewA1A548_Pd69e7TOJcOIBE06pOUVnjk},
  note         = {Accessed: July 17, 2025}
}

@book{krippendorff2018content,
  title={Content analysis: An introduction to its methodology},
  author={Krippendorff, Klaus},
  year={2018},
  publisher={Sage publications}
}

@article{seaman1999qualitative,
  title={Qualitative methods in empirical studies of software engineering},
  author={Seaman, Carolyn B.},
  journal={IEEE Transactions on software engineering},
  volume={25},
  number={4},
  pages={557--572},
  year={1999},
  publisher={IEEE}
}

@article{mchugh2012interrater,
  title={Interrater reliability: the kappa statistic},
  author={McHugh, Mary L},
  journal={Biochemia medica},
  volume={22},
  number={3},
  pages={276--282},
  year={2012},
  publisher={Hrvatsko dru{\v{s}}tvo za medicinsku biokemiju i laboratorijsku medicinu}
}

@article{freelon2010recal,
  title={ReCal: Intercoder reliability calculation as a web service},
  author={Freelon, Deen G},
  journal={International Journal of Internet Science},
  volume={5},
  number={1},
  pages={20--33},
  year={2010}
}

@article{cheng2021socially,
  title={Socially responsible ai algorithms: Issues, purposes, and challenges},
  author={Cheng, Lu and Varshney, Kush R and Liu, Huan},
  journal={Journal of Artificial Intelligence Research},
  volume={71},
  pages={1137--1181},
  year={2021}
}

@inproceedings{deng2023investigating,
  title={Investigating practices and opportunities for cross-functional collaboration around AI fairness in industry practice},
  author={Deng, Wesley Hanwen and Yildirim, Nur and Chang, Monica and Eslami, Motahhare and Holstein, Kenneth and Madaio, Michael},
  booktitle={Proceedings of the 2023 ACM Conference on Fairness, Accountability, and Transparency},
  pages={705--716},
  year={2023}
}

@inproceedings{holstein2019improving,
  title={Improving fairness in machine learning systems: What do industry practitioners need?},
  author={Holstein, Kenneth and Wortman Vaughan, Jennifer and Daum{\'e} III, Hal and Dudik, Miro and Wallach, Hanna},
  booktitle={Proceedings of the 2019 CHI conference on human factors in computing systems},
  pages={1--16},
  year={2019}
}

@article{van2020ethical,
  title={Ethical Implications of Artificial Intelligence: A Systematic Review of Bias, Fairness, and Accountability},
  author={Van Tuan, Nguyen and Quang, Tran Minh and others},
  journal={Artificial Intelligence and Machine Learning Review},
  volume={1},
  number={1},
  pages={1--7},
  year={2020}
}

@article{gunasekara2025systematic,
  title={A Systematic Review of Responsible Artificial Intelligence Principles and Practice},
  author={Gunasekara, Lakshitha and El-Haber, Nicole and Nagpal, Swati and Moraliyage, Harsha and Issadeen, Zafar and Manic, Milos and De Silva, Daswin},
  journal={Applied System Innovation},
  volume={8},
  number={4},
  pages={97},
  year={2025},
  publisher={MDPI}
}

@article{joshi2021ai,
  title={AI fairness via domain adaptation},
  author={Joshi, Neil and Burlina, Phil},
  journal={arXiv preprint arXiv:2104.01109},
  year={2021}
}

@article{yang2024survey,
  title={A survey of recent methods for addressing AI fairness and bias in biomedicine},
  author={Yang, Yifan and Lin, Mingquan and Zhao, Han and Peng, Yifan and Huang, Furong and Lu, Zhiyong},
  journal={Journal of Biomedical Informatics},
  volume={154},
  pages={104646},
  year={2024},
  publisher={Elsevier}
}

@article{singh2022fair,
  title={Fair CRISP-DM: Embedding fairness in machine learning (ML) development life cycle},
  author={Singh, Vivek and Singh, Anshuman and Joshi, Kailash},
  year={2022}
}

@article{sorathiya2024ethical,
  title={Ethical software requirements from user reviews: A systematic literature review},
  author={Sorathiya, Aakash and Ginde, Gouri},
  journal={arXiv preprint arXiv:2410.01833},
  year={2024}
}

@article{jakobssonrequirement,
  title={Requirement representation for safety-critical and fairness aware automotive perception systems},
  author={Jakobsson, Oskar and Rohacova, Zuzana},
year={2024}
}

@inproceedings{pushkarna2022data,
  title={Data cards: Purposeful and transparent dataset documentation for responsible ai},
  author={Pushkarna, Mahima and Zaldivar, Andrew and Kjartansson, Oddur},
  booktitle={Proceedings of the 2022 ACM Conference on Fairness, Accountability, and Transparency},
  pages={1776--1826},
  year={2022}
}

@article{rakova2021responsible,
  title={Where responsible AI meets reality: Practitioner perspectives on enablers for shifting organizational practices},
  author={Rakova, Bogdana and Yang, Jingying and Cramer, Henriette and Chowdhury, Rumman},
  journal={Proceedings of the ACM on Human-Computer Interaction},
  volume={5},
  number={CSCW1},
  pages={1--23},
  year={2021},
  publisher={ACM New York, NY, USA}
}

@article{ryan2023integrating,
  title={Integrating fairness in the software design process: An interview study with hci and ml experts},
  author={Ryan, Seamus and Nadal, Camille and Doherty, Gavin},
  journal={IEEE Access},
  volume={11},
  pages={29296--29313},
  year={2023},
  publisher={IEEE}
}

@article{pant2025navigating,
  title={Navigating fairness: practitioners’ understanding, challenges, and strategies in AI/ML development},
  author={Pant, Aastha and Hoda, Rashina and Tantithamthavorn, Chakkrit and Turhan, Burak},
  journal={Empirical Software Engineering},
  volume={30},
  number={3},
  pages={1--38},
  year={2025},
  publisher={Springer}
}

@article{ryanfairness,
  title={Fairness Challenges in the Design of Machine Learning Applications for Healthcare},
  author={Ryan, Seamus and Cai, Wanling and Bowman, Robert and Doherty, Gavin},
  journal={ACM Transactions on Computing for Healthcare},
  volume={6},
  number={4},
  pages={1--26},
  year={2025},
  publisher={ACM New York, NY}
}

@inproceedings{ferrara2024refair,
  title={Refair: Toward a context-aware recommender for fairness requirements engineering},
  author={Ferrara, Carmine and Casillo, Francesco and Gravino, Carmine and De Lucia, Andrea and Palomba, Fabio},
  booktitle={Proceedings of the IEEE/ACM 46th International Conference on Software Engineering},
  pages={1--12},
  year={2024}
}

@article{tang2025towards,
  title={Towards trustworthy AI-empowered real-time bidding for online advertisement auctioning},
  author={Tang, Xiaoli and Yu, Han},
  journal={ACM Computing Surveys},
  volume={57},
  number={6},
  pages={1--36},
  year={2025},
  publisher={ACM New York, NY}
}

@inproceedings{lu2022towards,
  title={Towards a roadmap on software engineering for responsible AI},
  author={Lu, Qinghua and Zhu, Liming and Xu, Xiwei and Whittle, Jon and Xing, Zhenchang},
  booktitle={Proceedings of the 1st International Conference on AI Engineering: software Engineering for AI},
  pages={101--112},
  year={2022}
}

@article{ferrara2024fairness,
  title={Fairness and bias in artificial intelligence: A brief survey of sources, impacts, and mitigation strategies},
  author={Ferrara, Emilio},
  journal={Sci},
  volume={6},
  number={1},
  pages={3},
  year={2024},
  publisher={Multidisciplinary Digital Publishing Institute}
}

@article{pham2025fairness,
  title={Fairness for machine learning software in education: A systematic mapping study},
  author={Pham, Nga and Ngoc, Hung Pham and Nguyen-Duc, Anh},
  journal={Journal of Systems and Software},
  volume={219},
  pages={112244},
  year={2025},
  publisher={Elsevier}
}

@article{ramadan2025towards,
  title={Towards Systematic Specification and Verification of Fairness Requirements: A Position Paper},
  author={Ramadan, Qusai and Ruohonen, Jukka and Tiwari, Abhishek and Alami, Adam and Boukhers, Zeyd},
  journal={arXiv preprint arXiv:2509.20387},
  year={2025}
}

@article{voria2024catalog,
  title={A catalog of fairness-aware practices in machine learning engineering},
  author={Voria, Gianmario and Sellitto, Giulia and Ferrara, Carmine and Abate, Francesco and De Lucia, Andrea and Ferrucci, Filomena and Catolino, Gemma and Palomba, Fabio},
  journal={arXiv preprint arXiv:2408.16683},
  year={2024}
}

@article{mccormack2024ethical,
  title={Ethical ai governance: Methods for evaluating trustworthy ai},
  author={McCormack, Louise and Bendechache, Malika},
  journal={arXiv preprint arXiv:2409.07473},
  year={2024}
}

@inproceedings{dehal2024exposing,
  title={Exposing algorithmic discrimination and its consequences in modern society: Insights from a scoping study},
  author={Dehal, Ramandeep Singh and Sharma, Mehak and de Souza Santos, Ronnie},
  booktitle={Proceedings of the 46th International Conference on Software Engineering: Software Engineering in Society},
  pages={69--73},
  year={2024}
}

@inproceedings{brun2018software,
  title={Software fairness},
  author={Brun, Yuriy and Meliou, Alexandra},
  booktitle={Proceedings of the 2018 26th ACM joint meeting on european software engineering conference and symposium on the foundations of software engineering},
  pages={754--759},
  year={2018}
}

@article{soremekun2022software,
  title={Software fairness: An analysis and survey},
  author={Soremekun, Ezekiel and Papadakis, Mike and Cordy, Maxime and Le Traon, Yves},
  journal={ACM Computing Surveys},
  year={2022},
  publisher={ACM New York, NY}
}

@incollection{garousi2020benefitting,
  title={Benefitting from the grey literature in software engineering research},
  author={Garousi, Vahid and Felderer, Michael and M{\"a}ntyl{\"a}, Mika V and Rainer, Austen},
  booktitle={Contemporary Empirical Methods in Software Engineering},
  pages={385--413},
  year={2020},
  publisher={Springer}
}

@article{fabris2025fairness,
  title={Fairness and bias in algorithmic hiring: A multidisciplinary survey},
  author={Fabris, Alessandro and Baranowska, Nina and Dennis, Matthew J and Graus, David and Hacker, Philipp and Saldivar, Jorge and Zuiderveen Borgesius, Frederik and Biega, Asia J},
  journal={ACM Transactions on Intelligent Systems and Technology},
  volume={16},
  number={1},
  pages={1--54},
  year={2025},
  publisher={ACM New York, NY}
}

@article{de2025software,
  title={Software Fairness Testing in Practice},
  author={de Souza Santos, Ronnie and de Morais Le{\c{c}}a, Matheus and Santos, Reydne and Magalhaes, Cleyton},
  journal={arXiv e-prints},
  pages={arXiv--2506},
  year={2025}
}

@article{ferrara2024fairnessinterview,
  title={Fairness-aware machine learning engineering: how far are we?},
  author={Ferrara, Carmine and Sellitto, Giulia and Ferrucci, Filomena and Palomba, Fabio and De Lucia, Andrea},
  journal={Empirical software engineering},
  volume={29},
  number={1},
  pages={9},
  year={2024},
  publisher={Springer}
}

@article{banks2025multiple,
  title={Multiple mini interviews vs traditional interviews: investigating racial and socioeconomic differences in interview processes},
  author={Banks, Pierre W and Hagedorn II, John C and Soybel, Alexandria and Coleman, Delayne Michelle and Rivera, Gabriel and Bhardwaj, Namita},
  journal={Advances in Medical Education and Practice},
  pages={157--163},
  year={2025},
  publisher={Taylor \& Francis}
}

\appendix

\end{document}